\documentclass[12pt,letterpaper]{JHEP}     
\usepackage{epsfig,feynmf}

\newfont{\frak}{eufm10 scaled 1200}

\newfont{\Bbb}{msbm10 scaled 1200}     
\newcommand{\mathbb}[1]{\mbox{\Bbb #1}}
\DeclareSymbolFont{AMSa}{U}{msa}{m}{n}
\DeclareSymbolFont{AMSb}{U}{msb}{m}{n}
\let\Box\relax
\DeclareMathSymbol{\Box}{\mathord}{AMSa}{"03}

\def\IZ{{\mathbb Z}}

\def\diag{\mbox{diag\,}}

\def \eqn#1#2{\begin{equation}#2\label{#1}\end{equation}} 

\def \Tr{\mbox{Tr\,}}

\def \lp{l_{planck}}

\def \ket#1{\left\vert#1\right\rangle}

\def \mth{M-theory}
\def\hacek{\accent20}                           
\def\calr{${\cal R\,}$}

\title{A nonsupersymmetric matrix orbifold}

\author{Tom Banks\\
  Department of Physics and Astronomy\\
  Rutgers University, Piscataway, NJ 08855-0849\\
E-mail: \email{banks@physics.rutgers.edu}}

\author{Lubo\hacek s Motl\\
  Department of Physics and Astronomy\\
  Rutgers University, Piscataway, NJ 08855-0849\\
E-mail: \email{motl@physics.rutgers.edu}}

\abstract{We construct the matrix description for a twisted version of the IIA
string theory on $S^1$ with fermions antiperiodic around a spatial circle.
The result is a  2+1-dimensional $U(N) \times U(N)$
nonsupersymmetric
Yang-Mills theory with fermionic matter transforming in the $({\bf 
N},{\bf \bar N})$. The
the two $U(N)$'s are exchanged if one goes around a twisted circle
of the worldvolume. Relations with Type 0 theories are explored and
we find Type 0 matrix string limits of our gauge theory.
We argue however that most of these results are falsified by the
absence of SUSY nonrenormalization theorems and that the models do
not in fact have a sensible Lorentz invariant space time interpretation.
}


\received{???????? ?st, 1999}
\accepted{???????? ?th, 1999}

\preprint{\hepth{9910164}\\RU-99-24\\HEP-UK-0009}
\begin{document}

\setlength{\unitlength}{1mm}
\begin{fmffile}{paperf1}   %


\section{The conformal field theory description}
%
Matrix Theory \cite{bfss} has been used to describe M-theory with
32 supercharges in 8,9,10 and 11 dimensions as well as various projections
of this theory.
In this paper we would like to study a nonsupersymmetric Matrix model 
in order to obtain a better understanding of SUSY breaking in string
theory. The problems of the model that we study show that SUSY breaking
leads to rather disastrous consequences.  However, we
point out in the conclusions that the restriction to the light cone
frame prevents us from abstracting completely clearcut lessons from this
exercise.  

Before proceeding, we note that after this paper was completed (but
before we had become convinced that the results were worth publishing)
another paper on Matrix models of nonsupersymmetric string theories
appeared \cite{jesus}.  We do not understand the
connection between the model presented there and the one we study.
Another recent paper on nonsupersymmetric compactifications, with
considerations related to ours is \cite{lust}.  Our results 
do not agree with the
suggestion of these authors that nonsupersymmetric compactifications
lead to Poincare invariant physics.
\vspace{3mm}

Let us start with the description of the conformal field theory that describes
the model we are interested in. We start with the Type IIA theory.
Compactification on a circle of radius $R$ can be described as modding
out the original theory by a symmetry isomorphic to $\IZ$, consisting of
the displacements by $2\pi k R$, $k\in\IZ$, in the chosen direction.
We can write those displacement as $\exp(2\pi i kR\hat p)$. A
``GSO-like'' projection by this operator
 now guarantees that the total momentum of a string
is a multiple of $1/R$. We also have to add ``twisted sectors'' where
the trip around the closed string is physically equivalent to any element
of the group that we divided by. Those sectors are wound
strings, $X(\sigma+2\pi)=X(\sigma)+2\pi Rw$, where $w\in\IZ$ is the
winding number.

\vspace{3mm}

Such a compactification preserves all 32 supercharges. We
will study a more complicated  model which breaks
the supersymmetry completely. The symmetry isomorphic to $\IZ$ will be
generated by (the direction of the circle is denoted by the index 2)
\eqn{generator}{G=\exp(2\pi i R\hat p_2)(-1)^F} 
where $(-1)^F$ counts the
spacetime statistics (or spin; in the Green-Schwarz formalism it is also
equivalent to the worldsheet spin). Because of this extra factor of
$(-1)^F$ the physics becomes very different. The fermionic fields of the
spacetime effective field theory become antiperiodic with the period 
$2\pi R_2$ while bosons are still periodic. Such a boundary condition of
course breaks supersymmetry completely because it is impossible to define
the (sign of the) supercharge everywhere. Those antiperiodic boundary
conditions for the fermions are the same as those used in finite temperature
calculations, with Euclidean time replaced by a spatial circle.
This compactification, introduced in \cite{rohm} is motivated by
Scherk-Schwarz compactifications of supergravity \cite{sesva}.
The physical spectrum is obtained by requiring $G\ket\psi=\ket\psi$ and in
the
twisted sectors corresponding to $G^w$, $w\in\IZ$, a trip around the
closed string is equivalent the shift
by $2\pi w R_2$ {\it times} $(-1)^{wF}$. Because $(-1)^{wF}$
for odd $w$ anticommutes with
fermions in the Green-Schwarz formalism, the fermions 
$\theta$ must be antiperiodic in the sectors with odd winding number.
Similarly, the GSO projection $G\ket\psi=\ket\psi$ now does not imply that
$p_2$ must be a multiple of $1/R_2$. Looking at 
(\ref{generator}) we see that there are two possibilities. Either
$(-1)^F$ is equal to $+1$ and $p_2=n/R_2$ {\it or}
$(-1)^F=-1$ and $p_2=(n+1/2)/R_2$. We have sectors with
$p_2R_2$ both integer or half-integer, but for integer $p_2R_2$ we
project out all the fermions and for half-integer $p_2R_2$ 
we project out all the bosonic states.

\vspace{3mm}

Thus we have four kinds of sectors; odd or even $w$ can be combined with
integer or half-integer $p_2R_2$.  For
even $w$ the boundary conditions are as in the untwisted theory but for
odd values of $w$ we must impose {\it antiperiodic} boundary
conditions for the Green-Schwarz fermions $\theta$.

Since the sectors with even values of $w$ are well-known (we just keep
only bosons or only fermions according to $p_2$),
we note only that in the sectors with odd values of
$w$ the ground state has $8\times (-1/24-1/48)=-1/2$ excitations both in
the left-moving and right-moving sector (the same as the ground state of a
NS-NS sector in the RNS formalism). In other words, the ground state is a
nondegenerate bosonic tachyon. We must also take into the account
the condition ``$L_0=\tilde L_0$'', more precisely ($N_L=N_R=0$ for the
ground state of even $w$ and $N_L=N_R=-1/2$ for odd $w$ and we define
$n=p_2R_2$) 
\eqn{levelmat}{N_L=N_R+nw.} 

Furthermore for integer $n$ we
must project all the fermions out of the spectrum. For even $w$ (which
means also even $nw$) this leaves us with the bosonic states of the
untwisted IIA theory. For odd $w$ (which implies integer $nw$) the
fermionic modes are half-integers and we see that due to
(\ref{levelmat}) the number of left-moving and right-moving fermionic 
excitations must be equal mod 2. Therefore the level matching condition
(\ref{levelmat}) automatically projects out all the fermionic states.

Similarly for half-integer $n$ we must get rid of all the bosonic states.
If $w$ is even, $nw$ is integer and odd and apart from (\ref{levelmat}) we
must also independently impose the condition $(-1)^F=-1$ and we get just
the fermionic part of the spectrum of the Type IIA string theory. However
for $w$ odd (which implies that $nw$ is half-integer and the fermions are
antiperiodic), we see a mismatch $1/2$ modulo $1$ between $N_L$ and $N_R$
in (\ref{levelmat}), so the bosons are projected out automatically as a
result of the level-matching condition (\ref{levelmat}).

\vspace{3mm}

In both cases we saw that the $(-1)^F$ projection was automatic in sectors
with odd values of $w$. It is a general property of 
orbifolds that in the twisted sectors where a trip around the closed
string is equivalent to the symmetry $g$, the GSO projection
$g\ket\psi=\ket\psi$ is a direct consequence of the level-matching
condition.

\vspace{3mm}

What about the tachyons? For even values of $w$ we have a part of the
spectrum of Type IIA strings, so there is no tachyon. However for odd
values of $w$, we can find a tachyon. Recall that
\eqn{masstach}{m^2=\frac{4}{\alpha'}N_L+
\left(\frac{n}{R_2}-\frac{wR_2}{\alpha'}\right)^2
=\frac{4}{\alpha'}N_R+\left(\frac{n}{R_2}+\frac{wR_2}{\alpha'}\right)^2.}

For $n=0$, $w=\pm 1$, we see that the ground level $N_L=N_R=-1/2$
is really tachyonic for $R_2^2/ \alpha' <2$. For sufficiently small 
radius $R_2$ there is a bosonic tachyon in the spectrum. If $R_2$ is
really small, also states with $\pm w=3,5,\dots$ (but always
$n=0$) may become tachyons.

However for $n=1/2$ (fermionic sector) and $w=1$ we see from
(\ref{levelmat}) that the lowest possible state has $N_R=-1/2$
and $N_L=0$ (one $\theta_{-1/2}$ left-moving excitation) which means that
$m^2$ expressed in (\ref{masstach}) is never negative. This means that the
tachyons can appear only in the bosonic spectrum (as scalars).

\vspace{3mm}

There are many interesting relations of such a nonsupersymmetric theory
with other theories of this kind. For example, by a Wick rotation we can
turn the twisted spatial circle into a time circle. The antiperiodic
boundary conditions for the fermionic field then describe a path integral
at finite temperature. The appearance of the tachyon in the spectrum 
for $R_2<\sqrt{2\alpha'}$ is related to the Hagedorn
phase transition. The infinite temperature, or zero $R_2$ limit of 
our model gives the Type~0 theories. 

The Type~0 theories (0A and 0B) are modifications of the Type~II theories
containing bosonic states only in ``diagonal sectors'' NS-NS and R-R; we
have also only one GSO projection counting the number of left-moving minus
right-moving fermionic excitations. The Type~II theories can be obtained
as an $\IZ_2$ orbifold (making separate projections on left-moving
fermions)
of the Type~0 theories in the R-NS formalism; the
difference between Type~IIA and Type~IIB theories is a sign of the
projection in the R-R sector; Type~IIA is in a sense Type~IIB with a
discrete torsion. Equivalently, we can also obtain Type~0 theories by
orbifolding Type~II theories but in the Green-Schwarz formalism: Type 0A
and 0B theories can be described in the Green-Schwarz formalism by the
same degrees of freedom as the corresponding Type II theories, but we must
include both PP and AA sector and perform the corresponding (diagonal)
GSO-projection.

\section{The matrix model}

In a first naive attempt to construct a model describing the Type~0A
theory, we would probably make a local orbifold (orbifolding in Matrix
theory was described in \cite{lmztwo}) of the original Matrix Theory,
corresponding to the $\IZ_2$ orbifold of the worldsheet theory in the
Green-Schwarz formalism. We would represent the operator $(-1)^F$ by a
gauge transformation e.g. $\sigma_3\otimes {\bf 1}$ and the bosonic
matrices would then be restricted to the block diagonal form, reducing the
original group $U(2N)$ into $U(N)\times U(N)$ with fermions in the
off-diagonal blocks i.e. transforming as $({\bf N},{\bf \bar N})$. However
the coordinates $X$ of the two blocks would suffer from an instability
forcing the eigenvalues of the two blocks to escape from each other: the
negative ground-state energy of the ``off-diagonal'' fermions is not
cancelled by a contribution of bosons and therefore the energy is
unbounded from below even for finite $N$.

Bergman and Gaberdiel however pointed out \cite{bg} that it is more
appropriate to think about Type 0A string theory as a Type IIA theory
orbifolded by a $\IZ_2$ group generated by the usual $(-1)^F$ {\it times}
the displacement by half of the circumference of the corresponding
M-theoretical circle. Of course, this displacement could not be seen
perturbatively. We are clearly led to the Scherk-Schwarz compactification
of M-theory. We will thus attempt to construct a more sophisticated
matrix model.  We will find a model which naively incorporates all of
the duality conjectures of Bergman and Gaberdiel and has Type 0A,B
and Rohm compactified Type IIA,B matrix string limits.  In the end, we
will find that many of our naive arguments are false, due to the absence
of SUSY nonrenormalization theorems, and that the model we construct 
does not have a Lorentz invariant large N limit.  We argue that this
implies that Scherk-Schwarz compactified \mth\ does not have a Lorentz
invariant vacuum. 

Let us start with a review of untwisted M-theory compactified on a circle.
The algorithm \cite{lmztwo} to mod out the BFSS model by a group of
physical symmetries $H$ is to enlarge the gauge group $U(N)$ and identify
elements of $H$ with some elements $g$ of the gauge group. It means that
the matrices $Y=X,\Pi,\theta$ are constrained to satisfy
\eqn{orbing}{h(Y)=g_hYg_h^{-1},\qquad g_h\in U(N),\quad h\in H.}

We wrote $gYg^{-1}$ because $Y$ transform in the adjoint representation
and the physical action of the symmetries on $Y$ is denoted $h(Y)$. To
obtain the $O(N)$ matrix model describing a single Ho\hacek rava-Witten
domain wall, we can set $g_h={\bf 1}$ and just postulate $Y$ to be
symmetric with respect to the symmetry (consisting of the reflection
of $X^1,\Pi^1$, multiplying spinors $\theta$ by $\gamma_1$ and transposing
all the matrices). Therefore $X^1$ and half $\theta$'s
become antisymmetric Hermitean matrices in the adjoint of $O(N)$ while
the other $X$'s and $\theta$'s become symmetric real matrices. The
naive matrix description of the heterotic strings on tori together with
the sectors and GSO-like projections on the heterotic matrix strings
was obtained in \cite{bamo}.

Compactification of $X^2$ on a circle with radius $R_2$
can be done in a similar
way. We just postulate the set of possible values of the $U(N)$
indices to be $\{1,2,\dots N\}\times (0,2\pi)_{circle}$ and
represent the physical symmetry
$\exp(2\pi i R_2\hat p)$ by the gauge transformation ${\bf 
1}\otimes\exp(i\sigma_2)$. Note that the matrices now have two discrete
and two continuous ``indices''. Postulating (\ref{orbing}) tells us
that the matrices must commute with any function of $\sigma$:
\eqn{gaufil}{X^2_{mn}(\sigma_2,\sigma'_2)=X^2_{mn}(\sigma_2)\delta(\sigma_2
-\sigma'_2)-i\delta'(\sigma_2-\sigma'_2)}
and similarly for the other matrices $X^i$ and $\theta$ (without the
$\delta'(\sigma_2-\sigma'_2)$ term). Now if we understand the summation
over
the sigma index as integration and ignore the factor $\delta(0)$
in the trace, the BFSS Hamiltonian becomes precisely the Hamiltonian of
SYM theory with $\sigma_2$ being an extra coordinate.

``Matrices'' of the form (\ref{gaufil}) can be also expressed in the
terms of the Fourier modes as done first by Taylor \cite{taylor}. 
The extra Fourier mode
indices replacing $\sigma_2,\sigma'_2$ are denoted $M,N$ and
(\ref{gaufil})
becomes
\eqn{gaudva}{(X^2_{M+1,N+1})_{mn}=(X^2_{M,N})_{mn}+2\pi R_2
\delta_{M,N}\delta_{mn}}
and similarly for the other matrices without the last term.

\vspace{3mm}

We will study a compactification of  M-theory on a twisted $T^2$ so we
will use two worldvolume coordinates $\sigma_1,\sigma_2$ to represent those
two circles.
What about the $(-1)^F$ twist which modifies the compactification
of $X^2$? Shifting both ends of the open strings
$M,N\to M+1,N+1$ as in (\ref{gaudva}) must be accompanied by $(-1)^F$
which commutes with bosons but anticommutes with spacetime fermions. In
the Green-Schwarz formalism it also anticommutes with the $\theta$'s. So the
structure of the bosonic matrices is unchanged and the
condition for $\theta$'s will be twisted:
\eqn{gautheta}{(\theta_{M+1,N+1})_{mn}=-(\theta_{M,N})_{mn}.}
In the continuous basis this is translated to
\eqn{gauc}{\theta_{mn}(\sigma_2,\sigma'_2)=\theta_{mn}(\sigma_2)\delta(\sigma_2
-\sigma'_2+\pi).}
This can be described by saying 
that $\theta$ has nonzero matrix elements between
opposite points of the $\sigma_2$ circle. We will often use this
``nonlocal'' interpretation of the resulting theory even though the theory
can be formulated as a conventional nonsupersymmetric gauge theory with 
fermionic matter, as we will show in a moment.

\subsection{The $U(N)\times U(N)$ formalism}

In order to get rid of the nonlocality, we must note that if we identify
the opposite points with $\sigma_2$ and $\sigma_2+\pi$, so that $\sigma_2$
lives on a circle of radius $\pi$, everything becomes local.  By halving
the circle, we double the set of bosonic fields. The two $U(N)$ groups at
points $\sigma_2$ and $\sigma_2+\pi$ are completely independent, so that
the gauge group becomes $U(N)\times U(N)$. We should also note that if we
change $\sigma_2$ by $\pi$, the two factors $U(N)$ exchange; this is an
important boundary condition.

The bosonic fields thus transform in the adjoint representation of
$U(N)\times U(N)$. What about the fermions $\theta$? We saw that the two
``matrix indices'' $\sigma_2$ and $\sigma'_2$ differ by $\pi$. One of them
is thus associated with the gauge group $U(N)$ at point $\sigma$, the
other with the $U(N)$ at point $\sigma+\pi$ which is the other $U(N)$
factor in the $U(N)\times U(N)$ formulation. In other words, $\theta$'s
transform as $({\bf N}, {\bf \bar N})$ under the $U(N)\times U(N)$. This
is a complex representation of (complex) dimension $N^2$ and the complex
conjugate $\theta^\dagger$s transform as $({\bf \bar N},{\bf N})$.  In the
old language, $\theta$ and $\theta^\dagger$ differed by $\pi$ in
$\sigma_2$ or in other words, they corresponded to the opposite
orientations of the arrow between $\sigma_2$ and $\sigma_2+\pi$. 
The number of real components is $2N^2$ (times the dimension of the spinor
16), the same as the dimension of the adjoint representation. This should
not surprise us since for $R_2\to\infty$ we expect that our
nonsupersymmetric model mimics the physics of the supersymmetric model.

We could also check that the commutation relations derived from the
``nonlocal'' orbifold formulation agree with the canonical commutation
relations of the $U(N)\times U(N)$ nonsupersymmetric gauge theory with the
matter in $({\bf N},{\bf \bar N})$. These theories are very similar to
the ``quiver'' theories of Moore and Douglas \cite{quiver}, but with a
peculiar boundary condition that exchanges the two $U(N)$ groups as we
go around the twisted circle.

\subsection{Actions for the local and nonlocal formulations}

We will be considering both descriptions. In one of them, the Yang-Mills
theory has gauge group $U(N)$ and is defined on a time coordinate
multiplied by a two-torus with
circumferences $1/R_1,1/R_2$ 
(instead of $2\pi$ employed in the previous section)
where $R_1,R_2$ are the radii of the
spacetime circles in Planck units\footnote{To simplify the presentation, we 
choose the convention for the numerical constants in various
dimensionful quantities to agree with these statements.} 
and the fermions are nonlocal degrees of freedom (arrows) pointing from
the point
$(\sigma_1,\sigma_2)$ to the point $(\sigma_1,\sigma_2+1/2R_2)$.
We will call this picture ``nonlocal''.

We will also sometimes use a ``local'' picture where the coordinate
$\sigma_2$ is wrapped twice and its circumference is only $1/2R_2$.
In the local picture, the gauge group is $U(N)\times U(N)$ and these two
factors exchange when we go around the $\sigma_2$ circle so that
the ``effective'' period is still equal to $1/R_2$:
\eqn{boundcond}{A^{\alpha}_{ij}(\sigma_0,\sigma_1,\sigma_2+1/2R_2)
=A^{1-\alpha}_{ij}(\sigma_0,\sigma_1,\sigma_2).}
Here $\alpha=0,1$ is an index distinguishing the two factors in $U(N)\times
U(N)$. We suppresed the
worldvolume vector index $\mu=0,1,2$.
Indices $i,j$ run from $1$ to $N$; here $i$ spans ${\bf N}$ and
$j$ belongs to ${\bf \bar N}$. 
Similar boundary conditions are imposed on the scalars $X$ which also
transform in the adjoint of $U(N)\times U(N)$. 
Both satisfy the usual hermiticity conditions.
Fermions $\theta$ (whose
spacetime transformation rules are the same as in the supersymmetric
theory) transform in $({\bf N},{\bf \bar N})$.
Writing them as $\theta_{ij}$, the index $i$ belongs to
${\bf N}$
of the first $U(N)$ and the index $j$ belongs to ${\bf \bar N}$ of the
second $U(N)$. In the same way, in
$(\theta^\dagger)_{ij}=(\theta_{ji})^\dagger$ the first index
$i$ belongs to ${\bf N}$ of the second $U(N)$ and the second index
$j$ belongs to ${\bf \bar N}$ of the first $U(N)$ so that 
$\Tr \theta^\dagger\theta=\theta^\dagger_{ij}\theta_{ji}$ is invariant.
The boundary condition for $\theta$s reads
\eqn{boundth}{\theta_{ij}(\sigma_0,\sigma_1,\sigma_2+1/2R_2)=
(\theta^\dagger)_{ij}(\sigma_0,\sigma_1,\sigma_2).}
Of course, $\theta$ matrices are complex, they do not obey a hermiticity
condition. The Lagrangian is $(i=1,\dots, 7)$
\eqn{lagnas}{{\cal L}=
\sum_{\alpha=0,1}\Tr\left[-\frac{1}{4}F_{(\alpha)}^{\mu\nu}F_{(\alpha),\mu\nu}
-\frac{1}{2}D_\mu X_{(\alpha)}^i D^\mu X_{(\alpha)}^i +\frac 14
[X^i_{(\alpha)},X^j_{(\alpha)}]^2
\right]}
$$+\Tr \left[i\theta^\dagger\gamma^i X^i_{(\alpha=0)} \theta
+i\theta\gamma^i X^i_{(\alpha=1)}\theta^\dagger 
+\theta^\dagger\gamma_\mu \partial^\mu \theta
+i\theta^\dagger\gamma_\mu A^\mu_{(\alpha=0)} \theta
+i\theta\gamma_\mu A^\mu_{(\alpha=1)}\theta^\dagger\right] $$
The trace always runs over $N\times N$ matrices.
We have put the dimensionful quantity $g_{YM}$ equal to one. The action is
simply
\eqn{akce}{{\cal A}=\int d\sigma^0\int_{0}^{1/R_1}d\sigma^1
\int_{0}^{1/(2R_2)}d\sigma^2 {\cal L}(\sigma^0,\sigma^1,\sigma^2).}
For the purposes of the calculations of Feynman diagrams it is also useful
to write the action in the nonlocal (nl) formulation of the theory.
In this formulation, the period of $\sigma^2$ is doubled and equal
to $1/R_2$. The fields can be identified as follows (the dependences
on $\sigma^0,\sigma^1$ and indices $\mu,i$ are suppressed):
\eqn{nelok}{X_{(\alpha)}(\sigma^2)=
X_{nl}(\sigma^2+\alpha/(2R_2)),\quad
A_{(\alpha)}(\sigma^2)=
A_{nl}(\sigma^2+\alpha/(2R_2)),\quad \alpha=0,1,}
\eqn{nelokdva}{\theta(\sigma^2)=\theta_{nl}(\sigma^2),\quad
\theta^\dagger(\sigma^2)=\theta^\dagger_{nl}(\sigma^2)=
\theta_{nl}(\sigma^2+1/(2R_2)),\quad
0\leq\sigma^2\leq \frac1{2R_2}.}
All the equalities are $N\times N$ matrix equalities.
In this nonlocal language the action can be written as
\eqn{nonlocakce}{{\cal A}=\int d\sigma^0\int_{0}^{1/R_1}d\sigma^1
\int_{0}^{1/R_2}d\sigma^2 {\cal L}_{nl}(\sigma^0,\sigma^1,\sigma^2)}
where (the subscript ``nl'' of all fields is suppressed)
\eqn{nonloclag}{{\cal L}_{nl}=
\Tr\left[-\frac{1}{4}F^{\mu\nu}F_{\mu\nu}
-\frac{1}{2}D_\mu X^i D^\mu X^i +\frac 14 [X^i,X^j]^2 \right]}
$$+\Tr \left[\theta^\dagger\gamma_\mu \partial^\mu \theta
+i\theta^\dagger(\gamma^i X^i(\sigma^2)+\gamma_\mu A^\mu(\sigma^2)) \theta
\right]$$
$$+\Tr\left[i\theta \left(\gamma^i
X^i(\sigma^2+\frac{1}{2R_2})+\gamma_\mu
A^\mu(\sigma^2+\frac{1}{2R_2})\right)
\theta^\dagger\right]$$
We denoted the $\sigma^2$ dependence in which one of the fermionic terms
is nonlocal.

\section{Alternative derivation and connection with Type 0 theories}

It has long been known \cite{seibwit,dixonhar} that the Type $0_{A,B}$
string theories in ten dimensions can be viewed as infinite temperature
limits of Type $II_{B,A}$ theories.  Rotating the Euclidean time to a
spacelike direction, this means that the zero radius limits of Rohm
compactifications are the Type 0 theories.  It is less well known (but, we
believe, known to many experts) that the finite radius Rohm
compactifications are compactifications of the Type 0 theories on dual
circles with a certain orbifold projection.  Indeed, both Type 0 theories
have two types of Ramond-Ramond fields which are related by a discrete
symmetry. This doubled number is a consequence of having one
GSO-projection only (the diagonal one). More precisely, the operator
\calr$=(-1)^{F_R}$ which counts the right-moving fermionic excitations has
eigenvalues $(+1)$ for half of the RR-fields and $(-1)$ for the other
half. Therefore $(-1)^{F_R}$ is a generator of a $\IZ_2$ symmetry that
exchanges the RR-fields in a basis rotated by 45 degrees, i.e.
$RR_{+1}+RR_{-1}$ with $RR_{+1}-RR_{-1}$ where
$RR_{\pm1}$ denotes the fields with $(-1)^{F_R}=\pm1$.

Twisted compactification of Type 0A or 0B with monodromy \calr\ i.e. the
orbifold of Type 0 string theory on a circle of circumference $2L$ by the
symmetry ${\cal R}\,\exp(iLp)$ gives a string model T-dual to the Rohm
compactified Type IIB or Type IIA string, respectively.  We will refer to
the twisted circle as the Scherk-Schwarz circle when describing it from
the Type II point of view and as the \calr circle from the Type 0 point of
view.

This T-duality is not hard to understand at the level of the string
spectrum. Because of the GSO projection, Scherk-Schwarz compactified Type
II theory contains bosonic states of integer momenta and fermionic states
of half-integer momenta (in appropriate units). The dual Type 0 string
theory initially had bosonic excitations only (in NS-NS and R-R sectors).
But because of the extra orbifold by ${\cal R}\,\exp(iLp)$, we obtain
also fermions in NS-R and R-NS twisted sectors with a half-integer
winding. This agrees with the assumption of T-duality. Apart from
T-duality between Type IIA/IIB on a Scherk-Schwarz circle and Type 0B/0A
on an \calr\ circle which we just mentioned, we should be aware of the
T-duality between Type 0A and Type 0B string theory on a usual circle.

Now consider the DLCQ of \mth\ compactified on a Scherk-Schwarz circle.
Using the logic of \cite{natiproof}, this is a zero coupling limit of Type
IIA string theory compactified on a Scherk-Schwarz circle of Planck size,
in the presence of $N$ D0-branes.  Using the T-duality adumbrated in the
previous paragraphs, this is weakly coupled Type 0B string theory on an
\calr\ circle in the presence of $N$ D-strings of the first kind; since
the \calr\ monodromy exchanges two types of D-strings, there must be an
equal number of D-strings of the other type. In other words, the D-strings
are compactified on a circle dual to the \mth\ Scherk-Schwarz circle, with
\calr\ twisted boundary conditions.

Now we can use the 
description of D-branes in Type 0 theories discovered by Bergman and
Gaberdiel \cite{bg}. As we have said, there are two types of D-strings in
Type 0B theory,
each of which has a bosonic $1+1$ dimensional gauge theory on its world
volume: open strings stretched between two like D-strings contain bosonic 
states only. These two types are related (exchanged) by the \calr\
symmetry. In the presence of closely spaced D-strings of both types,
there are additional fermionic degrees of freedom which transform in the 
$(N,\bar{M}) [\oplus (\bar{N},M)]$ of the $U(N) \times U(M)$ gauge group:
open strings stretched between two unlike D-strings contain fermions only.
These fermions are spacetime spinors. In the corresponding Seiberg
limit, all the closed string states (including tachyons) are decoupled
and in the corresponding DKPS energy scale only the massless open string
states survive.

The result is a $1+1$ dimensional $U(N) \times U(N)$ gauge theory with
fermions in the bifundamental and a boundary condition that exchanges the
two $U(N)$ groups as we go around the circle, a result of the \calr\
monodromy. This is the same gauge theory we arrived at by the orbifolding
procedure of the previous section. 

It is now easy to compactify an extra dimension on an ordinary circle and
obtain the $2+1$ dimensional gauge theory of the previous section as the
matrix description of Rohm compactification.  The double
T-duality in Seiberg's derivation can be done in two possible orders,
giving always the same result. In the next section, we will see that at
least formally we can rederive the various string theories as matrix
string limits of the gauge theory.  In particular, this will provide a
derivation of the Bergman-Gaberdiel duality relation between
Scherk-Schwarz compactification of \mth, and Type 0A strings.

\section{The matrix string limits}

\subsection{Rohm compactified Type IIA strings}

In the limit where the spacetime radius
$R_2$ goes to infinity, the radius of the worldvolume torus
$1/R_2$ goes to zero so that we also have $1/R_1\gg 1/R_2$. Therefore the
fields become effectively independent
of $\sigma^2$ up to a gauge transformation. Furthermore, because of the
boundary conditions exchanging the 
two $U(N)$'s, the expectation values of scalars in both $U(N)$'s must be
equal to each other (up to a gauge transformation). This can be also
seen in the nonlocal formulation: in the limit $R_2\to\infty$ the fields
must be constant (up to a gauge transformation) on the long circle of
circumference $1/R_2$.

Thus in this limit we can classify all the fields according to how they
transform under the $\sigma^2$ independent gauge symmetry $U(N)$. In the
nonlocal language this $U(N)$ is just a ``global'' (but $\sigma^1$
dependent) symmetry. In the $U(N)\times U(N)$ language this is the
diagonal symmetry $U(N)$. In both cases, we find that not only bosons but
also fermions (transforming originally in $({\bf N},{\bf \bar N})$)
transform in the adjoint (the same as ${\bf N}\otimes {\bf \bar N}$) of this
$U(N)$. There is only one set of fields: the $\sigma^2$ independence
causes the bosons in both $U(N)$'s to be equal and the complex matrices
$\theta$ to be Hermitean.

In the matrix string limit we expect to get a matrix description of the
Scherk-Schwarz compactification of Type~IIA strings on a long circle. The
appearance of the matrix strings (at a naive level) can be explained as
usual: most things work much like in the
supersymmetric matrix string theory \cite{lumodvv,bsdvv,dvv}.

In the nonlocal formulation, $U(N)$ gauge group is broken completely
down to a semidirect product of $U(1)^N$ and the Weyl group, $S_N$, 
of $U(N)$. Therefore the classical configurations around
which we expand are diagonalizable $N\times N$ matrices where the basis in
which they can be diagonalized can undergo a permutation $p\in S_N$
for $\sigma^1\to\sigma^1+1/R_1$:
\eqn{screwing}{X_i(\sigma^1)=U(\sigma^1)
\diag(x_i^1,x_i^2,\dots x_i^n)U^{-1},\qquad
U(\sigma^1+1/R_1)=U(\sigma^1)p.}

This is the mechanism of matrix strings \cite{lumodvv,bsdvv,dvv}. Every
permutation
$p$ can be decomposed into a product of cycles and each cycle of length
$k$ then effectively describe a ``long string'' with the longitudinal
momentum equal to $p^+=k/R^-$. For instance, a single cyclic permutation
of $k$ entries (written as a $k\times k$ matrix $p$) describes a single
string:
\eqn{permutace}{p=\left(\begin{array}{ccccc}
~\circ&~1&~\circ&\dots&\,\circ~\\
~\circ&~\circ&~1&\dots&\circ\\
\vdots&\vdots&\vdots&\ddots&1\\
~1&~\circ&~\circ&\dots&\circ\end{array}\right).}
The definition (\ref{screwing}) of $X_i$ creates effectively a string of
length $k$
(relatively to the circumference $1/R_1$). We can write
the eigenvalues as
\eqn{srouby}{x_i^m(\sigma^1)=x^{long}_i(\sigma^1+(m-1)/R_1),\qquad
m=1,2,\dots k}
where $x^{long}_i$ has period $k/R_1$. Assuming the $k/R_1$ periodicity of
$x^{long}_i$ we can show $1/R_1$ periodicity of the matrix 
(\ref{screwing}) with $p$ defined in (\ref{permutace}).

The matrix origin of the level-matching conditions was first explained
in \cite{dvv}: the residual symmetry $\IZ_k$ rotating the
``long'' string is a gauge symmetry and because the states must be
invariant under the gauge transformations, we find out that
$L_0-\tilde L_0$ must be a multiple of $k$ (the length of the string)
because the generator of $\IZ_k$ can be written as
$\exp(i(L_0-\tilde L_0)/k)$.
In the large $N$ limit such states are very heavy unless $L_0=\tilde L_0$
and we reproduce the usual level-matching conditions. In this limit the
discrete group $\IZ_k$ approximates the continuous group quite well.

\subsection{Dependence on the fluxes}

In the conformal field theory of the Rohm compactification,
sectors with odd or even winding numbers should have antiperiodic or
periodic spinors $\theta$, respectively.  We want to find the analog of
this statement in the Matrix formulation.

Let us put $w$ units of the magnetic flux in the nonlocal representation
of our theory. The corresponding potential can be taken to be
\eqn{magflux}{A_{\mu=1}=2\pi R_1R_2\sigma_2\frac wN, \quad
A_{\mu=2}=0}
Recall that the periods of $\sigma_1,\sigma_2$ are $1/R_1,1/R_2$.
In the local $U(N)\times U(N)$ formulation, the fields in
the region $0\leq \sigma_2 \leq 1/2R_2$ from (\ref{magflux})
define the block
of the first $U(N)$ and the region $1/2R_2 \leq \sigma_2 \leq 1/R_2$
defines the second $U(N)$. Note that for $\sigma_2\to\sigma_2+1/R_2$,
$\Tr A_{\mu=1}$ changes by $2\pi R_1 w$ which agrees with the
circumference of $X^1$.

Now if we substitute the background (\ref{magflux}) into (\ref{lagnas})
we see that the contributions of the form $\theta A\theta$ from the last
two terms give us a contribution coming from the 
difference $\sigma_2\to\sigma_2+1/2R_2$ which is equal to
\eqn{magff}{i\,\Tr[\theta^\dagger \gamma_1\theta]\frac{\pi w R_1}{N}.}
Such a term without derivatives would make the dynamics nonstandard.
However it is easy to get rid of it by a simple redefinition (we suppress
$\sigma_0,\sigma_2$ dependence)
\eqn{magredef}{\theta(\sigma_1) \to \theta(\sigma_1) 
\exp(i\sigma_1 R_1\pi w/N).}
Note that under $\sigma_1\to\sigma_1+N/R_1$ which corresponds to a loop
around a matrix string of length $N$, $\theta$ changes by a factor 
of $(-1)^w$. This confirms our expectations: in the sectors with an odd
magnetic flux (=winding number) the fermions $\theta$ are antiperiodic.

We might also wonder about the electric flux (=compact momentum $p_2$) in
the direction of $\sigma_2$. As we have explained in the beginning, this
flux should be allowed to take $1/2$ of the original quantum so that the
sectors with half-integer electric flux contain just fermions and the
usual sectors with integer electric flux contain bosons only, the other
being projected out by the GSO conditions in both cases. This behaviour
should be guaranteed ``by definition'': the operator $\exp(2\pi R_2\hat
p_2)(-1)^F$ is identified with a gauge transformation (namely
$\exp(2\pi R_2\sigma_2)\otimes {\bf 1}_{N\times N}$).

\vspace{2mm}

However it might seem a little strange that the together with the $\theta$
excitations one
must also change the electric flux; it might be useful to see the
origin of the sectors of
various electric flux ``microscopically''. We propose the following way to
think about this issue. The $\theta$ excitations in a compact space carry
charge $\pm$ with respect to groups $U(1)$ at the opposite points of
$\sigma_2$. The total charge vanishes therefore we do not have an
obstruction to excite $\theta$. However the charge does not vanish
locally, therefore we should accompany the excitation by an electric flux
tube running between $\sigma_2$ and $\sigma_2+1/2R_2$ in a chosen
direction (it is useful to think about it as a ``branch-cut'') and the
total electric flux induced by this excitation equals one half of the
quantum in the supersymmetric (untwisted) theory.

\vspace{2mm}

As a consequence of these observations, we see that our model contains 
the string field theory Hilbert space of the Rohm compactification in
the large $N$ limit.  At a very formal level, the dynamics of the model
in an appropriate limit of small radii and large Yang-Mills coupling,
appears to reduce to that of free Rohm strings.  However, this is not
necessarily a correct conclusion.  The analysis of the moduli space
Lagrangian is done at the classical level, but the apparent free string
limit corresponds to a strongly coupled YM theory.  In \cite{bsdvv} it
was emphasized that the derivation of Matrix string theory depends
crucially on the nonrenormalization theorem for the moduli space
Lagrangian.  We do not have such a theorem here and cannot truly derive
the free Rohm string theory from out model.  This is only the first of
many difficulties. 

A further important point is that the $U(1)$ gauge theory (or $U(1)\times
U(1)$ in the local formalism) leads to a free theory which is identical
to the conformal field theory in the matrix string limit
$1/R_1 \gg 1/R_2$. In particular we can see that the ground state in the
sectors with an odd magnetic flux has negative light cone energy, and
would have to be interpreted as a tachyon in a relativistic theory. 
Even if we
assumed the clustering property to be correct (in the next section we show
that this property is likely to be broken at the two-loop level), this tachyon
would lead to inconsistency in the large $N$ limit: it would be
energetically favoured for a configuration in the $U(N)$ theory to emit
the $N=1$ tachyonic
string -- and compensate the magnetic flux by the opposite value of the
flux in the remaining $U(N-1)$ theory. The energy of tachyon is of order
$-N^0$ which is negative and $N$ times bigger than the scale of energies
we would hope to 
study in the large $N$ limit (only states with energies of order $1/N$
admit a relativistic interpretation in the large $N$ limit).

To make this more clear: in order to establish the existence of a
relativistic large N limit we would have to find states with dispersion
relation ${{\bf p}^2 + m^2 \over N}$ in the model, as well as
multiparticle states corresponding to separated particles which scatter
in a manner consistent with relativity.  The observation of the previous 
paragraph shows that such states would generally be unstable to emission
of tachyons carrying the smallest unit of longitudinal
momentum\footnote{Note that in SUSY Matrix Theory the excitations along
directions where the gauge group is completely broken down to $U(1)$
factors have {\it higher} energy than the states with large longitudinal
momentum.}.  The only way to prevent this disaster is to lift the
moduli space.  However, once we imagine that the moduli space is lifted
it is unlikely that multiparticle states of any kind exist and the
model loses all possible spacetime interpretation.  We will investigate
the cluster property of our model below.  However, we first want to
investigate the Type 0 string limits of our model.  As above, we will
work in a purely classical manner and ignore the fact that the moduli
space is lifted by quantum corrections.

\subsection{Type 0 matrix strings}

The Rohm compactified IIA string is the formal limit of our $2+1$
dimensional gauge theory when the untwisted circle of the Yang Mills
torus is much
larger than the scale defined by the gauge coupling, while the twisted
circle is of order this scale or smaller.  We will now consider three
other limits.  The relation between the Yang-Mills parameters and 
the \mth\ parameters is
\eqn{ymcoup}{g_{YM}^2 = R/L_1 \ L_2}
\eqn{ymrad}{\Sigma_i = \lp^3 / R\ L_i \ .}
where $\Sigma_{1,2}$ is the untwisted (twisted) YM radius, $L_{1,2}$
are the corresponding \mth\ radii, and $R$ is the lightlike
compactification length.  In the Type 0 string limit, we want to take 
$L_2 \rightarrow 0$, with $\Sigma_2$ fixed (it is the string length
squared divided by $R$) and $L_1$ of order the string length).  The
latter restriction means that $g_{YM}^2\Sigma_1 $ is fixed.  
The limit is thus a $1+1$ dimensional gauge
theory on a fixed length twisted circle, with gauge coupling going to 
infinity.

Restricting ourselves to classical considerations, we are led to the
classical moduli space of this gauge theory.  The bosonic sector of the
moduli space consists of two sets of independent $N \times N$
diagonalizable  matrices.  However, in order to obtain configurations
which obey the twisted boundary conditions and have energy of order
$1/N$, one must consider only topological sectors in which the matrices
in the two gauge groups are identical.   Note however, that since the
bosonic variables are in the adjoint representation, they are not
affected by gauge transformations which are in the $U(1)$ subgroup.
This additional freedom becomes important when we consider the fermionic
variables.  The boundary conditions on these allow one other kind of
configuration with energy of order $1/N$: considering the fermions as
$N \times N$ matrices, we can allow configurations in which the
diagonal matrix elements come back to minus themselves (corresponding to
the gauge transformation $\pm (1, - 1)$ in $U(N)\times U(N)$) after a
cycle with length of order $N$.  The resulting low energy degrees of
freedom are fermion fields on the ``long string'' with either periodic
or antiperiodic boundary conditions.  The gauge fields are vector like
so in terms of left and right moving fields we get only the PP and AA
combinations of boundary conditions.
The $O(8)$ chirality of   
the fermions is correlated to the world sheet chirality as in IIA 
matrix string theory.  One also obtains a GSO projection on these
fermionic degrees of freedom by imposing the gauge projection
corresponding to the $(1, -1)$  transformation.  The resulting model is
thus seen to be the Type 0A string theory, written in light cone
Green-Schwarz variables.   Remembering that the $1+1$ twisted gauge
theory was the matrix description of Scherk-Schwarz compactification of
M-theory, we recognize that we have derived the conjecture of Bergmann and
Gaberdiel. 

To obtain the 0B matrix string limit and the T-duality (on an untwisted
circle) between the two Type 0 theories, we simply follow the results of
one of the present authors and Seiberg \cite{bsdvv} and first take the 
strongly coupled Yang Mills limit by going to the classical moduli space
and performing a $2+1$ dimensional duality transformation.  
This corresponds to both directions of the Yang Mills torus being much
larger than the Yang Mills scale.  We then do a
dimensional reduction to a $1+1$ dimensional theory to describe
the 0B and IIB string limits.  In the former, the twisted circle is
taken much larger than the untwisted one, while their relative sizes are
reversed in the latter limit.
After the duality transformation and dimensional
reduction the manipulations are identical to those reported above.

A serious gap in the argument is the absence of $2+1$ superconformal
invariance.  In \cite{bsdvv} this was the crucial fact that enabled one to
show that the interacting Type IIB theory was Lorentz invariant.  Here
that argument fails.  We view this as an indication that the
spacetime picture derived from free Type 0 string theory is misleading.
We will discuss this further below.  Indeed, in the next subsection we 
show that the cluster property which is at the heart of the derivation
of spacetime from Matrix Theory fails to hold in our model.

\subsection{Breakdown of the cluster property}

The easiest way to derive the Feynman rules is to use the nonlocal formulation
(\ref{nonloclag}). It looks similar to a local Lagrangian except that
the gauge field in the last term (i.e. the  whole third line) is taken from
$\sigma^2+1/2R^2$. In the
Feynman diagrams the propagators have (worldvolume) momenta in the lattice
corresponding to the compactification, i.e. $P_1,P_2$ are
multiples of $2\pi R_1$ or $2\pi R_2$ respectively. The last term
in (\ref{nonloclag}) gives us a vertex with two fermions and one gauge
boson and the corresponding Feynman vertex contains a factor
$(-1)^{P^2/2\pi R_2}$.

In order to determine the cluster properties of our theory, we must
calculate the effective action along the flat directions in the
classical moduli space of the gauge theory.  We will concentrate on 
a single direction in which (in the nonlocal formulation) the gauge
group is broken to $U(N_1) \times U(N_2)$.  That is, we calculate
two body forces, rather than general $k$ body interactions.  There is a
subtlety in this calculation which has to do with our lack of knowledge 
of the spectrum of this nonsupersymmetric theory.

In general, one may question the validity of the Born-Oppenheimer 
approximation for the flat directions because the individual nonabelian
gauge groups appear to give rise to infrared divergences in perturbation
theory.  In the SUSY version of Matrix Theory this problem is resolved
by the (folk) theorem that the general $U(N)$ theory (compactified on a
torus) has threshold bound states.  These correspond to wave functions
normalizable along the flat directions and should cut off the infrared
divergences.  In our SUSY violating model, we do not know the relevant
theorems.  

The most conservative way to interpret our calculation is to take $N_1 =
N_2 = 1$ in the $U(2)$ version of the model.  If one finds an attractive
two body force then it is reasonable to imagine that in fact the general
$U(N)$ theory has a normalizable ground state, thus justifying the 
Born-Oppenheimer approximation in the general case.

So let us proceed to calculate the potential in the $U(2)$ case, and let
$R$ be the field which represents the separation between two excitations
of the $U(1)$ model.    
>From the point of view of $2+1$ dimensional field theory, $R$ is a
scalar field, with mass dimension $1/2$.  It is related to the distance
measured in M-theory by powers of the eleven dimensional Planck scale.
At large $R$, the charged fields of the $U(2)$ model are very heavy.
To integrate them out we must understand the UV physics of the model.
The formulation in terms of a $U(2)\times U(2)$ theory with peculiar
boundary conditions shows us that the UV divergences are of the same
degree as those of the SUSY model, though some of the SUSY cancellations
do not occur, as we will see below.  
Ultraviolet physics is thus dominated by the
fixed point at vanishing Yang-Mills coupling and we can compute the
large $R$ expansion of the effective action by perturbation theory.

\vspace{0.5cm}

$$
\qquad 
 \begin{fmfgraph*}(15,15)
  \fmfforce{(0,.5h)}{v1}
  \fmfforce{(w,.5h)}{v2} 
  \fmf{boson,left}{v1,v2}
  \fmf{boson,label=$ $,l.d=.1w,l.s=right,left}{v2,v1}
 \end{fmfgraph*}
\qquad
 \begin{fmfgraph*}(15,15)
  \fmfforce{(0,.5h)}{v1} 
  \fmfforce{(w,.5h)}{v2}
  \fmf{dashes,left}{v1,v2}
  \fmf{dashes,label=$ $,l.d=.1w,l.s=right,left}{v2,v1}
 \end{fmfgraph*}
\qquad
 \begin{fmfgraph*}(15,15)
  \fmfforce{(0,.5h)}{v1} 
  \fmfforce{(w,.5h)}{v2}
  \fmf{fermion,left}{v1,v2}
  \fmf{fermion,label=$ $,l.d=.1w,l.s=right,left}{v2,v1}
 \end{fmfgraph*}\qquad
$$ 
\vspace{-1.9cm}

{\quad\qquad\qquad\qquad\qquad\qquad $p$
\qquad\qquad\quad $p$
\qquad\qquad\quad $p$}

\vspace{1cm}

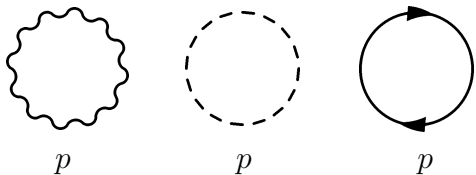
{\qquad\qquad\qquad {\bf Fig.1:} One-loop diagrams.}

\vspace{1cm}

$$
 \begin{fmfgraph*}(22,22) 
  \fmfforce{(0,.5h)}{v1}
  \fmfforce{(w,.5h)}{v2}
  \fmf{fermion,left}{v1,v2}
  \fmf{fermion,l.d=.1w,l.s=right,left}{v2,v1}
  \fmf{boson,label=$ $}{v1,v2}
  \fmfdot{v1,v2}
 \end{fmfgraph*}
\qquad\qquad\qquad
 \begin{fmfgraph*}(22,22)
  \fmfforce{(0,.5h)}{v1}
  \fmfforce{(w,.5h)}{v2}
  \fmf{fermion,left}{v1,v2}
  \fmf{fermion,l.d=.1w,l.s=right,left}{v2,v1}
  \fmf{dashes,label=$ $}{v1,v2}
  \fmfdot{v1,v2}
 \end{fmfgraph*}
$$
\vspace{-2cm}

\mbox{$\qquad\qquad\qquad\quad V_1\qquad\quad p\qquad\quad
V'_2$$\qquad\quad
V_1\qquad\quad p\qquad\quad V'_2$}

\vspace{0.8cm}

\nopagebreak
{\qquad\qquad\qquad {\bf Fig.2:}
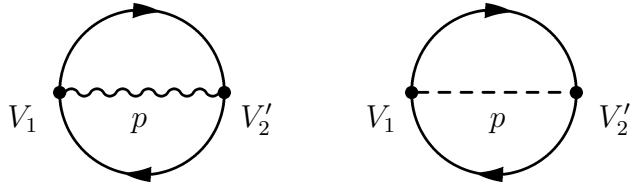
Non-vanishing two-loop contributions to the}\newline
{${~}_{~}$\qquad\qquad\qquad\qquad\qquad\quad effective potential. Here
$V'_2\equiv (\epsilon(p)-1)V_2$.}

\vspace{5mm}

The one loop contribution, $Fig. 1$,
to the effective potential vanishes because it
is identical to that in the SUSY model.  The only difference between the
models in the nonlocal formulation is the peculiar vertex described above. 
The leading contribution comes from two loops and is of order $g_{YM}^2$
(the squared coupling has dimensions of mass).  It comes only from the 
diagrams containing fermion lines shown in $Fig. 2$.  These diagrams
should be evaluated in the nonlocal model and then their value in the
SUSY model should be subtracted.  The rest of the two loop 
diagrams in the model are the same as the SUSY case and they cancel (for
time independent $R$) against the SUSY values of the diagrams shown.
Taking $R$ very large in the diagrams is, by dimensional analysis,
equivalent to taking the volume large, and the potential is extensive in
the volume in the large volume limit.  The massive particles in the
loops have masses of order $g_{YM} R$ and this quantity is kept fixed in
the loop expansion.

Our Lagrangian has two gauge boson fermion vertices $V_1 + \epsilon (p) V_2$.  
$p$ is the momentum.
In the SUSY theory, these are the two terms in the commutator. In our
Lagrangian, the
first is identical to that in the SUSY theory while the second differs
from it by the sign $\epsilon (p)$, which is negative 
for odd values of the loop momentum around the twisted
circle.  Schematically then, the nonvanishing two loop contribution 
has the form
\eqn{twoloop}{\langle(V_1 + \epsilon (p) V_2)^2\rangle - \langle (V_1 +
V_2)^2 \rangle}
This can be rewritten as
\eqn{twolooptwo}{2 \langle (\epsilon (p) - 1) V_1 V_2 \rangle}

The resulting loop integral is quadratically ultraviolet divergent.
The leading divergence is independent of $R$, but there are subleading
terms of order $g_{YM}^2 \Lambda \left\vert g_{YM} R\right\vert$ and
$g_{YM}^2
{\ln}(\Lambda)
(g_YM R)^2$.  Corrections higher order in the Yang-Mills coupling, as well
as
those coming from finite volume of the Yang-Mills torus, are subleading
both in $R$ and $\Lambda$. Thus, the leading order
contribution to the potential is either confining, or gives a disastrous
runaway to large $R$. Which of these is the correct behavior is
determined by our choice of subtractions.  It would seem absurd to
choose the renormalized coefficient of $R^2$ to be negative, and obtain
a Hamiltonian unbounded from below.  
If that is the case, then a confining potential prevents excitations
from separating from each other in the would-be transverse spacetime. 
In other words, the theory does not have a spacetime interpretation at
all, let alone a relativistically invariant one.

We also see that the fear expressed in the previous chapter that all
excitations will decay into tachyons of minimal longitudinal momentum
was ill founded.  Instead it would appear that the entire system will
form a single clump in transverse space.  
The $U(1)$ part of the theory decouples, so we can give this clump
transverse momentum and obtain an energy spectrum
\eqn{enspec}{P^- \sim {R {\bf P}^2 \over N} + \Delta}
where $\Delta$ is the ground state energy of our nonlocal $SU(N)$
Yang-Mills theory.  

If $\Delta$ were to turn out positive and of order $1/N$, this
dispersion relation would look like that of a massive relativistic
particle.  We might be tempted to say that the system looked like a
single black hole propagating in an asymptotically flat spacetime.
The stability of the black hole would be explained if its mass were
within the Planck regime.
This interpretation 
does not appear to be consistent, for semiclassical analysis of
such a system indicates that it has excitations corresponding to
asymptotic gravitons propagating in the black hole background.  Our
result about the lifting of the moduli space precludes the existence of
such excitations.  

Since the vacuum energy is divergent, the positivity of $\Delta$ is a
matter of choice.  However, large $N$ analysis suggests that it scales 
like a positive power of $N$, so we have another reason that 
the black hole interpretation does not
seem viable.  
\section{Conclusions}

What are we to make of all these disasters?  We believe that our work is
solid evidence for the absence of a Lorentz invariant vacuum of M-theory
based on the Rohm compactification.  The Rohm strings are certainly 
degrees of freedom of our matrix model, even if we cannot derive the
(apparently meaningless because of the tachyon and unbounded effective
potential) string perturbation expansion from it (as a consequence of
the absence of a nonrenormalization theorem).

However, naive physical intuition based on the string perturbation
series, suggest that if there is a stable solution corresponding to
the Rohm model, it is not Lorentz invariant.  While we cannot trust the
perturbative calculations in detail, they at least imply that the vacuum
energy of the system is negative at its (hypothetical) stable minimum.
(We remind the reader that even at large radius, before the tachyon
appears, the potential calculated by Rohm is negative and the system
wants to flow to smaller radius).
Perhaps there is a nonsupersymmetric Anti-DeSitter solution of M-theory
to which the Rohm model ``flows''.  There are many problems
with such an interpretation, since it involves changing asymptotic
boundary conditions in a generally covariant theory.  Normally one would 
imagine that M-theory with two different sets of asymptotic boundary
conditions breaks up into two different quantum mechanical systems which
simply do not talk to each other.  The finite energy states with one set
of boundary conditions simply have no overlap with the finite energy
states of another (the definition of energy is completely different).

As an aside we note that an 
extremely interesting question arises for systems (unlike the Rohm
compactification) which have a {\it metastable} Minkowski vacuum.  In
the semiclassical approximation \cite{cdl} one can sometimes find
instantons which represent tunneling of a Minkowski vacuum into a
``bubble of Anti-DeSitter space''.  Does this idea make any
sense in a fully quantum mechanical theory, particularly if one believes 
in the holographic principle?  Coleman and De Luccia argue that the
system inside the AdS bubble is unstable to recollapse and interpret
this as a disaster of cosmic proportions: Minkowski 
space fills up with bubbles, 
expanding at the speed of light, the interior of each of which becomes
singular in finite proper time.  It is hard to imagine how such a
scenario could be described in a holographic framework\footnote{This
should not be taken simply as an indication that a holographic
description of asymptotically flat spacetimes is somehow sick.  The
Coleman De Luccia instanton also exists in an asymptotically AdS
framework, with two negative energy vacua.  If the higher energy
state has very small vacuum energy the semiclassical analysis is
practically unchanged.  So, if the Coleman DeLuccia phenomenon really
exists in M-theory we should be able to find a framework for studying
it within AdS/CFT.}.   

At any rate, it is clear that the fate of the Rohm compactification
depends crucially on a change in vacuum expectation values.  In this
sense one might argue that our attempt to study it in light cone frame 
was ``doomed from the start''.  It is a notorious defect of the light
cone approach that finding the correct vacuum is extremely difficult.
It involves understanding and cancelling the large $N$ divergences of
the limiting DLCQ, by changing parameters in the light cone Hamiltonian.
If the correct vacuum is a finite distance away in field space from the
naive vacuum from which one constructs the original DLCQ Hamiltonian,
this may simply mean that the true Hamiltonian has little resemblance to
the one from which one starts.  If our physical arguments above are
a good guide, the problem may be even more severe.  The correct vacuum
may not even have a light cone frame Hamiltonian formulation.

We confess to having jumped in to the technical details of our
construction before thinking through the physical arguments above.
Nonetheless, we feel that our failure is a useful reminder that
gravitational physics is very different from quantum field theory, and
an indication of the extreme delicacy of SUSY breaking in M-theory.

\acknowledgments

This work was supported in part by the DOE under grant
number DE-FG02-96ER40559.$\!\!$


\end{fmffile}
\end{document}